\documentclass[preprint,review]{elsarticle}
\usepackage{amsmath,amsfonts,amssymb}
\usepackage{verbatim}
\usepackage{algpseudocode}
\usepackage{graphicx}
\usepackage{textcomp} 
\usepackage{multirow} 
\usepackage{epstopdf}
\usepackage{color}
\usepackage{algorithm}
\usepackage{threeparttable}
\usepackage[table,xcdraw]{xcolor}
\usepackage{subfigure}
\usepackage{changepage}
\usepackage{bm}
\usepackage{diagbox}
\usepackage{float}

\newtheorem{remark}{Remark}

\newcommand{\fr}{\mathrm{r}}
\newcommand{\ft}{\mathrm{t}}
\newcommand{\fT}{\mathrm{T}}

\newcommand{\fH}{\mathrm{H}}

\newcommand{\bt}{\mathbf{b}}

\newcommand{\et}{\mathbf{e}}
\newcommand{\hht}{\mathbf{h}}

\newcommand{\st}{\mathbf{s}}
\newcommand{\ut}{\mathbf{u}}
\newcommand{\vt}{\mathbf{v}}
\newcommand{\wt}{\mathbf{w}}

\newcommand{\mta}{\boldsymbol{\theta}}

\newcommand{\mTa}{\mathbf{\Theta}}

\newcommand{\mvG}{\boldsymbol{\varGamma}}

\newcommand{\At}{\mathbf{A}}

\newcommand{\GGt}{\mathbf{G}}
\newcommand{\Ht}{\mathbf{H}}
\newcommand{\Wt}{\mathbf{W}}
\newcommand{\Pt}{\mathbf{P}}
\newcommand{\Qt}{\mathbf{Q}}

\newcommand{\bbC}{\mathbb{C}}

\usepackage{lineno}

\journal{Knowledge-Based Systems}

\begin{document}
	\begin{frontmatter}
	\title{Heuristic Solution to Joint Deployment and Beamforming Design for STAR-RIS Aided Networks}
	\author[a,b]{Bai Yan}
	\ead{yanb@sustech.edu.cn}
	
	\author[b]{Qi Zhao}
	\ead{zhaoq@sustech.edu.cn}
	
	\author[b]{Jin Zhang\corref{cor1}}
	\ead{zhangj4@sustech.edu.cn}
 \cortext[cor1]{Corresponding author}	 
	
	\author[c]{J. Andrew Zhang}
 \ead{Andrew.Zhang@uts.edu.au}

	\address[a]{Research Institute of Trustworthy Autonomous Systems, Southern University of Science and Technology, Shenzhen 518055, China}
	
	\address[b]{Department of Computer Science and Engineering, Southern University of Science and Technology, Shenzhen 518055, China}
	
	\address[c]{Global Big Data Technologies Centre, University of Technology Sydney, NSW 2007, Australia}

\begin{abstract}
This paper tackles the deployment challenges of Simultaneous Transmitting and Reflecting Reconfigurable Intelligent Surface (STAR-RIS) in communication systems. Unlike existing works that use fixed deployment setups or solely optimize the location, this paper emphasizes the joint optimization of the \textit{location} and \textit{orientation} of STAR-RIS. This enables searching across all user grouping possibilities and fully boosting the system's performance. We consider a sum rate maximization problem with joint optimization and hybrid beamforming design. An \textit{offline} heuristic solution is proposed for the problem, developed based on differential evolution and semi-definite programming methods. In particular, a point-point representation is proposed for characterizing and exploiting the user-grouping. A balanced grouping method is designed to achieve a desired user grouping with low complexity. Numerical results demonstrate the substantial performance gains achievable through optimal deployment design.
\end{abstract}

\begin{keyword}
Simultaneous transmitting and reflecting reconfigurable intelligent surface (STAR-RIS) \sep deployment \sep heuristic \sep evolutionary algorithm
\end{keyword}

\end{frontmatter}

\newpage

\section{Introduction}
Reconfigurable intelligent surface (RIS) has been recently proposed as a revolutionizing technology to customize the propagation environment of communication systems \cite{huang2022combining,yuan2020intelligent,wu2019towards}. A RIS is a planar passive radio structure composed of several reconfigurable passive elements. Each RIS element can independently adjust its phase shift and amplitude on the incident signal \cite{9360709}. Consequently, a directional beam is generated by the collaborative efforts of RIS elements to improve the received signals or eliminate inter/intra-cell interference. Compared with traditional active relaying, RIS is more energy-efficient, as it does not amplify the noise or require active radio-frequency chains \cite{di2020smart}. 

Various types of RIS have emerged. The most common types include reflecting-only and transmitting-only RISs, which only create a half-space smart radio environment\cite{basharat2021reconfigurable,basar2021present,YAN2022109725}. To overcome this drawback, a frontier type, simultaneous transmitting and reflecting reconfigurable intelligent surface (STAR-RIS) \cite{xu2021simultaneously,zhang2022intelligent}, has been proposed. With STAR-RIS, the incident signals are simultaneously transmitted and reflected towards users on both sides of STAR-RIS. Full-space coverage and new degree-of-freedom (DoFs) can be achieved to boost the system performance \cite{liu2021star}. STAR-RIS protocol includes energy splitting, model switching and time switching modes \cite{9570143}. Under these protocols, the \textit{hybrid beamforming} design, i.e., passive beamforming at the STAR-RIS and active beamforming at the base station (BS), was investigated in various communications systems to achieve different objectives \cite{9570143,cai2022joint,lu2021max}.


However, most STAR-RIS works employ fixed deployment setups \cite{lu2021max,zuo2022joint,wu2022resource,zhang2022secrecy}, restricting the system performance. Several works have started exploring the potential benefits of optimizing the STAR-RIS location. Some focus on identifying a desired location but with fixed \textit{user grouping} assumption \cite{gao2022joint,wang2023average,xiao2024star}. This indicates that users are classified to either the transmission or reflection side of STAR-RIS, and this classification does not change during location optimization. Literature \cite{pan2023joint} delves into deploying multiple STAR-RISs within a finite set of candidate locations. Others \cite{zhang2022joint,su2023joint,aung2023aerial} consider limited user-grouping configurations. 

Though these efforts in location optimization achieve performance improvements, they overlook optimizing the orientation. The location and orientation jointly determine the user grouping, further influencing the hybrid beamforming design. The interdependence contains up to $K^2$ user-grouping possibilities in a $K$-user scenario. Solely optimizing the location explores a subset of potential user groupings, thus failing to fully exploit the benefits of user grouping for optimal system performance. Moreover, the uncertainty in user grouping incurs varying types of decision variables (transmission or reflection type), exacerbating the complexity of the deployment and hybrid beamforming design. 

In this paper, we focus on jointly optimizing the \textit{location} and \textit{orientation} of STAR-RIS. Considering a downlink STAR-RIS-assisted multi-user multiple-input single-output (MU-MISO) communication system, we aim to maximize the sum rate by jointly optimizing the STAR-RIS deployment and hybrid beamforming, subject to the transmit power constraint at BS and the quality-of-service (QoS) constraint for each user. Given a STAR-RIS location, an orientation change alters the channel directivity with or without the user grouping. By contrast, the variations in user grouping yield substantial performance discrepancies. It is essential to thoroughly characterize the user grouping to fully exploit its benefits. Meanwhile, the traditional gradient-based methods would fail to handle the uncertainty in user grouping. To address these issues, we propose an \textit{offline} heuristic solution, called Differential Evolution with Balanced Grouping (DEBG), developed by exploiting differential evolution \footnote{We choose DE due to its excellent simulation results compared to other evolutionary algorithms. Please see details in the supplement file.} (DE) and semi-definite programming (SDP) methods. Our contributions are:
\begin{itemize}
	\item We formulate a sum rate maximization problem for joint optimization of STAR-RIS deployment and active beamforming at BS and passive beamforming at STAR-RIS, subject to the transmit power constraint at BS and the QoS constraint for each user. Unlike previous works, we, for the first time, jointly optimize the orientation and location of STAR-RIS to achieve quasi-optimal system performance.  
	
	\item We propose a point-point representation to characterize the user grouping, where one point is STAR-RIS and the other is a chosen user. A line crossing the two points decides the user grouping. This representation is advantageous to fully exploiting the benefits of user grouping, providing a solid basis for better problem-solving. 
	
	\item We propose a balanced grouping method to choose a user as the other point of the point-point presentation. In this case, a desired user grouping can be achieved based on the line across the STAR-RIS and this user. It can heuristically enhance the system performance with low complexity. 
	
	\item We validate DEBG's performance via simulations. Results demonstrate that DEBG achieves significant performance improvement compared to other deployment counterparts. Deploying the STAR-RIS near the users with symmetric deployment strategies is preferable.
	
\end{itemize} 

The rest of this paper is organized as follows. Sections II and III provide the system model and problem formulation. Section IV presents the DEBG approach. Section V gives simulation results and discussions. Finally, section VI concludes the paper.

\textbf{\textit{Notations}}: $\mathbb{R}^{M}$ and $\mathbb{C}^{N}$ denote the space of ${M}$ real-valued and ${N}$ complex-valued vector, respectively. $(\cdot)^\fT$ and $(\cdot)^\fH$ denote transpose and conjugate transpose of a vector or matrix, respectively. \text{Rank($\At$)} and \text{Tr}({$\At$}) denote the rank and trace of matrix $\At$, respectively. $\At\succeq0$ means that $\At$ is a positive semidefinite matrix. $\mathcal{CN}(0, \sigma^2)$ indicates a Gaussian distribution with zero mean and variance $\sigma^2$. 

\section{System Model}
\begin{figure}[t]
	\centering
	\includegraphics[width=0.37\textwidth]{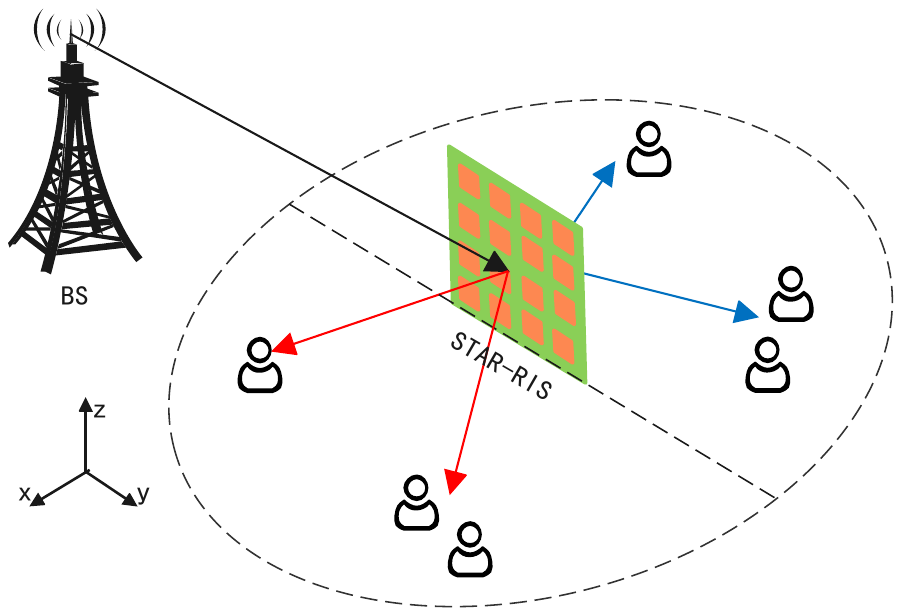}
	\caption{A downlink STAR-RIS assisted MU-MISO communication network.}
	\label{fig-system}
\end{figure}

This paper considers a narrow-band downlink STAR-RIS assisted MU-MISO communication system, as shown in Fig. \ref{fig-system}. A BS with $N_a$ antennas transmits signals to $K$ single-antenna users. All the users are assumed to be static or low-mobility \cite{gao2022joint,9570143}, i.e., a quasi-static scenario. The direct links between the BS and users are assumed to be blocked by the same-altitude buildings or other obstacles. Therefore, we hang a high-altitude STAR-RIS to provide non-line-of-sight links to enhance the link quality. The BS transmits its signals to the users with the transmission- or reflection- functionality of the STAR-RIS. We assume that all the perfect channel state information (CSI) is available at the BS \cite{9774942, 9739715}. The channel acquisition methods are out of the scope of our work, which will not be discussed in this paper. In a 3D coordinate system, we define the locations of the BS's centre antenna, STAR-RIS's centre and user $k$ as $\bt=(x_b,y_b,z_b)^\fT$, $\st=(x_s,y_s,z_s)^\fT$, and $\ut_k=(x_k,y_k,z_k)^\fT$, where 
\begin{equation}\label{eq-STAR-location-constraint}
		\setlength{\abovedisplayskip}{3pt}
	\setlength{\belowdisplayskip}{3pt}
		\st\in\mathcal{S}=\{(x_s,y_s,z_s)^\fT|x_s\in[x_{min},x_{max}], y_s\in[y_{min},y_{max}], z_s\in[z_{min},z_{max}]\},
\end{equation}
where $[x_{min},x_{max}]$, $[y_{min},y_{max}]$, and $[z_{min},z_{max}]$  denote the ranges along the $x-$, $y-$ and $z-$ axes. The orientations of BS antennas are fixed, denoted as Euler angles $\mathbf{o}_B$. The STAR-RIS plane is deployed perpendicular to the ground (fixed pitch and yaw angles) and has a 1D orientation (roll angle) around the z-axis. Thus, we denote the STAR-RIS's orientations as Euler angles $\mathbf{o}_S=(0, 0, \phi)$. $\mathbf{o}_i$ ($i\in\{B,S\}$) can also be expressed by the rotation matrix $\mathbf{T}_i$, where $\mathbf{T}_i^\fT\mathbf{T}_i=\mathbf{I}^{3}$, $\text{det}(\mathbf{T}_i)=1$\cite{zheng2023jrcup}. 

In the RIS's local coordinate system, we denote the elevation angle-of-departure (AoD) and the azimuth AoD towards user $k$ as $\varphi^{{SU}_k}$ and $\psi^{{SU}_k}$. The elevation angle-of-arrival (AoA) and the azimuth AoA from BS are denoted as $\varphi^{SB}$ and $\psi^{SB}$. They follow	
	\begin{equation}\label{eq-angles}
		\begin{aligned} 
			\varphi^\mathrm{SU_k}&=\arctan2\left([\mathbf{T}_S^\fT(\ut_k-\st)]_2, [\mathbf{T}_S^\fT(\ut_k-\st)]_1\right),\\
			\psi^\mathrm{SU_k}&=\arcsin\left([\mathbf{T}_S^\fT(\ut_k-\st)]_3/\|\ut_k-\st\|_2\right),
		\end{aligned}		
	\end{equation}		
In the BS's local coordinate system, the elevation AoD and the azimuth AoD towards the STAR-RIS are denoted as $\varphi^{BS}$ and $\psi^{BS}$. A relationship similar to \eqref{eq-angles} holds for the pairs $\{\varphi^{BS}, \psi^{BS}\}$ and $\{\varphi^{SB}, \psi^{SB}\}$.

\subsection{STAR-RIS Model}
We consider the energy splitting protocol \cite{9570143} for STAR-RIS, where the incident signal is divided into a transmission and a reflected signal by the STAR-RIS. We choose this protocol because of its easier hardware implementation than the time switching protocol and better performance than the model switching protocol \cite{9570143}. To characterize the energy splitting protocol, the transmission- and reflection-beamforming vectors are defined as 
\begin{equation}
		\setlength{\abovedisplayskip}{3pt}
	\setlength{\belowdisplayskip}{3pt}
	\left\{
	\begin{aligned}    	
		\mta_\ft=[\sqrt{\beta^\ft_1}e^{j\theta^\ft_1},\sqrt{\beta^\ft_2}e^{j\theta^\ft_2},...,\sqrt{\beta^\ft_M}e^{j\theta^\ft_M}]^\fT,\\ \mta_\fr=[\sqrt{\beta^\fr_1}e^{j\theta^\fr_1},\sqrt{\beta^\fr_2}e^{j\theta^\fr_2},...,\sqrt{\beta^\fr_M}e^{j\theta^\fr_M}]^\fT,
	\end{aligned}
	\right.\\
	\label{eq-STAR-beamforming vectors}
\end{equation}  
where $M$ is the number of STAR-RIS elements; $\sqrt{\beta^\ft_m}$, $\sqrt{\beta^\fr_m}\in[0,1]$ and $\theta^\ft_m$, $\theta^\fr_m\in[0,2\pi)$ with $m\in\mathcal{M}=\{1,...,M\}$ denote the amplitude and phase shift imposed on the incident signal by the $m$-th STAR-RIS element, respectively. Note that each transmission- and reflection- phase shift can be adjusted independently. However, due to the law of energy conservation, the transmission- and reflection- amplitude of each STAR-RIS element are coupled, which satisfies $\beta^\ft_m+\beta^\fr_m=1$. Then, the passive beamforming diagonal matrix between the STAR-RIS and user $k$ is denoted as 
\begin{equation}\label{eq-STAR coefficient diagonal matrix}
		\setlength{\abovedisplayskip}{3pt}
	\setlength{\belowdisplayskip}{3pt}
	\mTa_k=\left\{
	\begin{aligned}    	
		&\text{diag}(\mta_\ft),\ \text{if user}\ k\ \text{is at transmission half-space}, \\ 
		&\text{diag}(\mta_\fr),\ \text{if user}\ k\ \text{is at reflection half-space},
	\end{aligned} \right. 	
\end{equation}  
where $k\in\mathcal{K}=\{1,...,K\}$.

\subsection{Transmission Model}
Let $\GGt\in\bbC^{M\times N_a}$ and $\vt_k\in\bbC^{M\times1}$ represent the channels BS-to-STAR and STAR-to-user-$k$, which are modeled as Rician fading channels \cite{mu2021joint}
\begin{equation}
			\setlength{\abovedisplayskip}{3pt}
	\setlength{\belowdisplayskip}{3pt}
	\begin{aligned}
		\GGt&=\sqrt{\frac{\beta_{BS}}{1+\beta_{BS}}}\GGt^\mathrm{LoS}+\sqrt{\frac{1}{1+\beta_{BS}}}\GGt^\mathrm{NLoS}, \\
		\vt_k&=\sqrt{\frac{\beta_{SU_k}}{1+\beta_{SU_k}}}\vt_k^\mathrm{LoS}+\sqrt{\frac{1}{1+\beta_{SU_k}}}\vt_k^\mathrm{NLoS},		
	\end{aligned}
	\label{eq-Rician-fading-channels}
\end{equation}
where $\beta_{BS}$ and $\beta_{SU_k}$ denote the Rician factor of the BS-to-STAR and STAR-to-user-$k$ links, respectively. $\GGt^\mathrm{NLoS}$ and $\vt_k^\mathrm{NLoS}$ are the random non-line-of-sight (NLoS) components modeled as Rayleigh fading, distributed as $\mathcal{CN}(0,1)$. $\GGt^\mathrm{LoS}=\breve{\mathbf{a}}(\varphi^\mathrm{BS},\psi^\mathrm{BS})\mathbf{a}(\varphi^\mathrm{SB}, \psi^\mathrm{SB})^\fT$ and $\vt_k^\mathrm{LoS}=\mathbf{a}(\varphi^\mathrm{SU_k}, \psi^\mathrm{SU_k})$ denote the deterministic line-of-sight (LoS) components of the BS-to-STAR and STAR-to-user-$k$ channels. $\breve{\mathbf{a}}(\varphi^\mathrm{BS},\psi^\mathrm{BS})\in\mathcal{C}^{N_a}$ is the array response vector of the BS, while $\mathbf{a}(\varphi^\mathrm{SB}, \psi^\mathrm{SB})\in\mathcal{C}^{M}$ and $\mathbf{a}(\varphi^\mathrm{SU_k}, \psi^\mathrm{SU_k})\in\mathcal{C}^{M}$ are the array response vectors of the STAR-RIS. They are defined as \cite{zheng2023jrcup}
		\begin{equation}
					\setlength{\abovedisplayskip}{3pt}
			\setlength{\belowdisplayskip}{3pt}
		\begin{aligned}
		\relax [\breve{\mathbf{a}}(\varphi^\mathrm{BS}, \psi^\mathrm{BS})]_{n_a}&=e^{j\frac{2\pi}{\lambda}\mathbf{p}_{B,n_a}^\fT\mathbf{t}(\varphi^\mathrm{BS}, \psi^\mathrm{BS})},\\
		[\mathbf{a}(\varphi^\mathrm{SB}, \psi^\mathrm{SB})]_m&=e^{j\frac{2\pi}{\lambda}\mathbf{p}_{S,m}^\fT\mathbf{t}(\varphi^\mathrm{SB}, \psi^\mathrm{SB})},\\
		[\mathbf{a}(\varphi^\mathrm{SU_k}, \psi^\mathrm{SU_k})]_m&=e^{j\frac{2\pi}{\lambda}\mathbf{p}_{S,m}^\fT\mathbf{t}(\varphi^\mathrm{SU_k}, \psi^\mathrm{SU_k})},
		\end{aligned}
		\label{eq-array vector1}
	\end{equation}
where $\mathbf{p}_{B,{N_a}}$ and $\mathbf{p}_{S,m}$	are positions of the $n_a$-th BS element and $m$-th STAR-RIS element in their local coordinate systems. $\lambda$ is the signal wavelength. $\mathbf{t}(\varphi,\psi)\triangleq[\cos(\varphi)\cos(\psi), \sin(\varphi)\cos(\psi),\sin(\psi)]^\fT$ is the direction vector.

Furthermore, the path loss of the BS-to-STAR-to-$u_k$ channel is given by 
\begin{equation}
	\setlength{\abovedisplayskip}{3pt}
	\setlength{\belowdisplayskip}{3pt}
	L_k=\frac{\rho_0}{d_{BS}^{\alpha}}\frac{\rho_0}{d_{SU_k}^{\alpha}}
	\label{eq-path-loss}
\end{equation}
where $\rho_0$ is the path loss at a reference distance of 1 meter, $d_{BS}=\|\st-\bt\|$ and $d_{SU_k}=\|\st-\ut_k\|$ denote the distances of the BS-to-STAR and STAR-to-user-$k$, respectively. $\alpha$ is the corresponding path loss exponent. 

Therefore, the received signal at user $k$ can be expressed as
\begin{equation}
	\setlength{\abovedisplayskip}{3pt}
	\setlength{\belowdisplayskip}{3pt}
	\mathtt{y}_k=\underbrace{\hht_k\wt_k e_k}_{\text{Desired signal}} + \underbrace{\sum_{j\neq k, j\in\mathcal{K}}\hht_k\wt_j e_j}_{\text{Interference}} + \underbrace{n_k}_{\text{Noise}},
	\label{eq-yk}
\end{equation}
where $\hht_k=\sqrt{L_k}\vt_k^\fH\mTa_k\GGt$ is the cascaded channel of the BS-STAR-RIS-user-$k$ link, $\wt_k$ and $e_k$ denote the active beamforming vector and transmitted symbol to user $k$ at the BS, where $e_k\sim\mathcal{CN}(0, 1)$. $n_k\sim\mathcal{CN}(0, \sigma_0^2)$ is the white Gaussian noise at user $k$. The user $k$ treats all the signals from other users as interference. Therefore, the communication rate of user $k$ is
\begin{equation}
	\setlength{\abovedisplayskip}{3pt}
	\setlength{\belowdisplayskip}{3pt}
	R_k=\log_2\left(1+\frac{|\hht_k\wt_k|^2}{\sum_{j\neq k, j\in\mathcal{K}}|\hht_k\wt_j|^2+\sigma_0^2}\right).
	\label{eq-SR}
\end{equation}

\section{Proposed Problem Formulation}
We aim to maximize the sum rate by jointly designing the location and orientation of STAR-RIS and hybrid beamforming, subject to the transmit power constraint of BS and QoS constraint for each user. Unlike previous works, the STAR-RIS orientation is alterable and optimized, such that all potential user-grouping scenarios can be exploited to fully enhance the system performance. 

Specifically, we formulate a joint deployment and hybrid beamforming problem for sum rate maximization
\begin{subequations}\label{pb-SR}
		\setlength{\abovedisplayskip}{3pt}
	\setlength{\belowdisplayskip}{3pt}
	\begin{align}
		\max_{\st, \phi, \mvG}\
		&\sum_{k=1}^{K}R_k, \tag{\ref{pb-SR}{a}}\label{pb-SRa}\\ 
		s.t.\ &\st\in\mathcal{S},\ \phi\in[0,2\pi), \tag{\ref{pb-SR}{b}}\label{pb-SRb} \\
		&\sum_{k=1}^{K}\|\wt_k\|^2\leqslant P_{max}, \tag{\ref{pb-SR}{c}}\label{pb-SRc}\\
		& R_k\geq R_{min}, \tag{\ref{pb-SR}{d}}\label{pb-SRd}\\
		&\sqrt{\beta^\ft_m}, \sqrt{\beta^\fr_m}\in[0,1],\ \beta^\ft_m+\beta^\fr_m=1,\ \theta^\ft_m, \theta^\fr_m\in[0,2\pi),\ \forall m\in\mathcal{M}, \tag{\ref{pb-SR}{e}}\label{pb-SRe}\\
		&\mTa_k=\begin{cases}    	
			\text{diag}(\mta_\ft), &if\ \text{BS and $u_k$ are at different}\ \text{sides of}\ f(x), \\
			\text{diag}(\mta_\fr), &if\ \text{BS and $u_k$ are at the same}\ \text{side of}\ f(x),
		\end{cases} \tag{\ref{pb-SR}{f}}\label{pb-SRf} 	
	\end{align}
\end{subequations}	
where the STAR-RIS's location and orientation are $\st$ and $\phi$; $\mvG=(\wt_k, \mta_\ft, \mta_\fr)$ is the hybrid beamforming, with $\wt_k, \mta_\ft$ and $\mta_\fr$ being the active beamforming to user $k$ at BS, the transmission- and reflection- beamforming vectors of STAR-RIS, respectively. (\ref{pb-SRa}) aims at maximizing the sum rate of all users. (\ref{pb-SRb}) is the constraint for the STAR-RIS deployment\footnote{More realistic deployment scenarios with obstacles will be investigated in future.}. (\ref{pb-SRc}) is the total transmit power constraint of the BS, where $P_{max}$ is the maximum transmit power. (\ref{pb-SRd}) is the QoS constraint for each user, where $R_{min}$ is the minimum rate requirement. (\ref{pb-SRe}) is the constraint for the amplitudes and phase shifts of STAR-RIS. (\ref{pb-SRf}) is the user grouping constraint, where $f(x)$ is a straight line that traverses the STAR-RIS with angle $\phi$, named as ``boundary line". We obtain $f(x)=\tan\phi\times(x-x_s)+y_s$ according to geometry. This line decides whether a user is served by the transmission- or reflection- beamforming vectors\footnote{We forbid the BS and users to be on the boundary line since the user grouping and channel modelling cannot be estimated. It is easy to implement in practice.}. 

For ease of problem-solving, we first address the deployment and hybrid beamforming problem \eqref{pb-SR} \textit{offline}, leveraging the deterministic LOS components to achieve a favourable deployment scheme. Since the STAR-RIS is deployed in high places to avoid signal blockage, the deterministic LOS components are expected to be the dominant factor. After deploying the STAR-RIS based on the favourable scheme, the instantaneous CSI of the transmission- and reflection- users are obtained consecutively by the CSI estimation techniques for RISs \cite{9133156,9130088}. The hybrid beamforming is refined online with accurate CSI to further enhance performance. This paper only focuses on offline optimization as similar optimization strategies can be applied to hybrid beamforming. 

Problem (\ref{pb-SR}) is very challenging to solve since the location and orientation of STAR-RIS have a complex and profound effect on the system performance. The location affects the path loss of the BS-STAR-RIS-user $k$ channel, i.e., $L_k$. The location and orientation co-decide the channel directivity. Notably, the \textit{uncertainty} in user grouping \eqref{pb-SRf} incurs varying types of decision variables ($\mta_\ft$ or $\mta_\fr$), exacerbating the design complexity of deployment and hybrid beamforming. 

\section{Proposed Heuristic Solution}
An orientation change alters the channel directivity with or without the user grouping. By contrast, the variation of user grouping would bring great performance variation. It is necessary to characterize the user grouping to fully exploit its benefits. Therefore, we propose a novel point-point representation to define user grouping. Traditional gradient-based methods would fail to handle the uncertainty issue in user grouping. For this, we propose a heuristic solution, DEBG, developed on DE and SDP methods. A major innovation in DEBG is the balanced grouping method, which strategically balances user QoS on both sides of the STAR-RIS, yielding high-quality user groupings with minimal complexity.

\begin{figure}[t]
	\centering
	\includegraphics[width=0.37\textwidth]{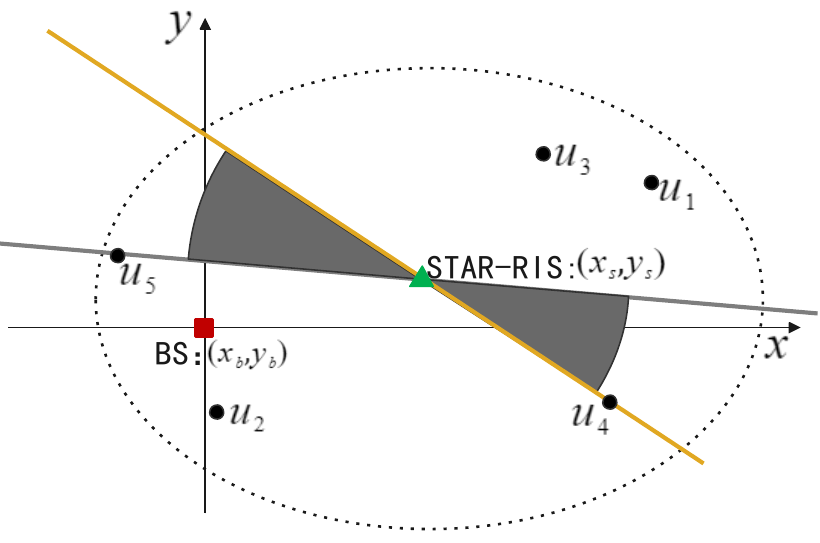}
	\caption{User-grouping based on the point-point representation (top view).}
	\label{fig-point-point}
\end{figure}

\subsection{Point-Point Representation for User-Grouping}
We propose a point-point representation to characterize the user grouping, where the two points are the locations of STAR-RIS and a specific user. Fig. \ref{fig-point-point} gives a visual example. Imagine a straight line traversing the STAR-RIS and running parallel to the x-axis. Rotate this line counterclockwise with the STAR-RIS as the centre, and arrange the users swept by this rotation as $u_1,...u_K$. The line that intersects the STAR-RIS and a chosen user $u_c$ with a tiny counterclockwise rotation angle offset (e.g., $0.001^\circ$), is termed the \textit{boundary line} (shown in yellow), which determines the user grouping. We choose these two points because i) the STAR-RIS location remarkably affects the path loss of the BS-STAR-RIS-$u_k$ channel, and ii) the variation of $u_c$ under a given STAR location results in different user groupings, which is advantageous for exploring the optimal user grouping.

Mathematically, we denote the point-point representation as $(\st, c)$, where $\st$ is the STAR-RIS's location, and $c\in\mathcal{K}$ is named as boundary-user index. With $(\st, c)$, Problem (\ref{pb-SR}) is recast as
\begin{subequations}\label{pb-SR2}
	\setlength{\abovedisplayskip}{3pt}
	\setlength{\belowdisplayskip}{3pt}
	\begin{align}
		\max_{\st, c, \phi, \mvG}\
		&\sum_{k=1}^{K}R_k, \tag{\ref{pb-SR2}{a}}\label{pb-SR2a}\\ 
		s.t.\ & \st\in\mathcal{S},\ c\in\mathcal{K},\ \tag{\ref{pb-SR2}{b}}\label{pb-SR2b} \\
		&\phi\in[\phi_{(c)},\phi_{(c+1)})+0.001,\ \tag{\ref{pb-SR2}{c}}\label{pb-SR2c} \\
		&(\ref{pb-SRc}), (\ref{pb-SRd}), (\ref{pb-SRe}), (\ref{pb-SRf}). \tag{\ref{pb-SR2}{d}}\label{pb-SR2d} 	
	\end{align}
\end{subequations}

\subsection{Framework}
\begin{figure}[t]
	\centering
	\includegraphics[width=0.75\textwidth]{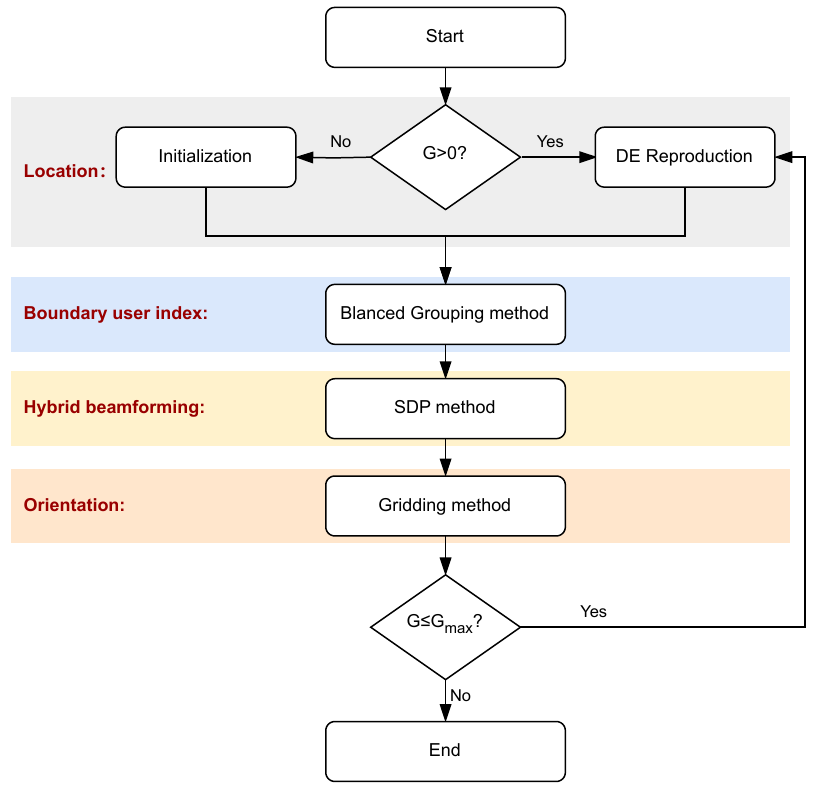}
	\caption{DEBG workflow.}
	\label{fig-workflow}
\end{figure}

To solve Problem (\ref{pb-SR2}), we propose a heuristic DEBG solution (Fig. \ref{fig-workflow}). It is a population-based search approach. Let $G$ and $G^{max}$ be the current and maximum generation respectively. For the ongoing population, the STAR-RIS locations $\st$s, boundary user indexes $c$s, STAR-RIS orientations $\phi$s and hybrid beamformings $\mvG$s are updated sequentially. The core components of DEBG include i) DE reproduction, which yields the offspring solutions' locations; ii) balanced grouping, which finds a matching boundary user for each individual to achieve a high-quality user grouping. iii) SDP method, which calculates the hybrid beamforming, and ix) gridding method, for orientation refinement.

\subsubsection{DE Reproduction and DE Selection}
As a classical evolutionary algorithm, DE efficiently handles various gradient-free and mixed variable-type optimisation problems \cite{storn1997differential,lin2019differential,wei2021penalty}. Meanwhile, we conduct simulations to compare the performance of DE and other evolutionary algorithms in the supplementary file. The simulation results demonstrate the efficacy of DE. Therefore, we employ the mutation and crossover operators of DE to update the locations of the STAR-RIS. The DE selection operator selects high-quality solutions to enter the next generation. 

\textbf{Mutation:} Among different mutation strategies \cite{price2013differential}, this paper uses the DE/best/1 strategy due to its fast convergence speed. With this strategy, a mutation vector $\vt_i$ is produced by
\begin{equation} \label{eq-mutation}
		\setlength{\abovedisplayskip}{3pt}
	\setlength{\belowdisplayskip}{3pt}
	\boldsymbol{\upsilon}_i=\st_{best}+F\times (\st_{a_1}-\st_{a_2}),
\end{equation}
where $\st_{best}$ is the STAR-RIS location of the individual with the best fitness. $a_1$ and $a_2$ are the indexes of two individuals randomly selected from the population. $F$ is a positive real parameter called the ``amplification factor'', which controls the amplification of the difference vectors. 

\textbf{Crossover:} After mutation, DE performs a binomial crossover operator that exchanges some exponents from $\st_i$ and $\boldsymbol{\upsilon}_i$ to form a trial vector $\widetilde{\st}_{i}$:
\begin{equation} 
		\setlength{\abovedisplayskip}{3pt}
	\setlength{\belowdisplayskip}{3pt}
	\widetilde{\st}_{i,j}=\left\{
	\begin{aligned}
		&\st_{i,j},\ \text{if rand}\leq Cr\ \text{or}\ j=j_{rand}, \\
		&\boldsymbol{\upsilon}_{i,j},\ \text{otherwise},   
	\end{aligned} \right.
	\label{eq-crossover}
\end{equation}
where $j_{rand}$ is an integer uniformly selected from \{1,2,3\} to make sure that at least one dimension of $\widetilde{\st}_i$ comes from $\boldsymbol{\upsilon}_i$. $Cr\in[0,1]$ is the crossover rate, which decides the proportion of $\widetilde{\st}_{i}$ inherited from the mutation vector $\boldsymbol{\upsilon}_i$.

\textbf{Selection:} Note that the balanced grouping, SDP and gridding method (will be detailed in following sections) are implemented after crossover to produce the offspring $\Qt^G$. Then, the selection operator selects the better individual from each pair of individuals ($\Pt_i^G$ and $\Qt_i^G$) for entering the next generation, where $\Pt_i^G$ and $\Qt_i^G$ are the $i$-th parent and offspring solution at the $G$-th generation. The superiority of feasible solution technique \cite{7277038} is employed to select individuals, in which: i) for two feasible solutions, the one with better fitness (i.e., higher sum rate) is selected; ii) a feasible solution is always better than an infeasible one; iii) for two infeasible solutions, the one with smaller constraint violation (i.e., the QoS constraint violation) is preferred. 

\begin{algorithm}[t]
	\caption{\textit{Balanced\_Grouping}} 
	\begin{algorithmic}[1]
		\State Calculate the differential distances $\{D_c\}$s by Eq. (\ref{eq-d-both sides}); 
		\State Compute the selection probability of $u_c$ being the boundary user: $p_c=D_c/(\sum_{c=1}^{K}D_{c})$;
		\State Sort $\{p_c\}$s in increasing order and obtain a list of sorted values $\{p_\text{s}\}$s, where the ranking of $p_c$ is defined as $\hat{c}$;
		\State Compute the cumulative selection probability of user $c$: $q_c=\sum_{i=1}^{\hat{c}}p_\text{s}(i)$;	
		\State Pick a random value $\tau$ uniformly from $[0,1]$;
		\State set $i=1$;
		\While{$q_i<\tau$}
		\State $i=i+1$;
		\EndWhile	
		\State Find the user with ranking $i$, and its index is the boundary user's index.
	\end{algorithmic}
	\label{al-RW-grouping}
\end{algorithm}

\subsubsection{Balanced Grouping} \label{sec-RW grouping}
Given a STAR-RIS's location, the optimal boundary user can be obtained by exhaustively searching over all possible boundary users and choosing the best candidate boundary user. However, searching over every boundary user requires solving a hybrid beamforming problem with prohibitive complexity. To tackle this issue, we propose a fast balanced grouping method based on the STAR-RIS-user link distances and the number of users on both sides of the STAR-RIS. This strategy does not need to perform hybrid beamforming or exhaustive search, which balances convergence and complexity. 

The intuition of balanced grouping is as follows. We divide the sum rate of all users into two parts: the sum rate of transmission users and that of reflection users. If the two parts are both maximized, the sum rate of all users would be maximized. But how to balance the two parts? To handle this issue, we consider using the STAR-to-user distances and the number of users to design a balance measure. It is based on thoughts on Eqs. (\ref{eq-path-loss})-(\ref{eq-SR}):

1) If the distance of STAR-to-user-$k$ (i.e., $d_{SU_k}$) is small, the path loss $L_k$ is small, the received signals of user $k$ are strong, hence user $k$'s rate is high. 

2) To maximize the sum rate of all users, the sum rate of users on both sides of the STAR-RIS should both be maximized. One intuitive solution is, the average distance of the STAR-RIS-to-transmission-users and that of the STAR-RIS-to-reflection-users should both be minimized.

3) However, if we only minimize the average distance of STAR-RIS-to-transmission/reflection-users, many close-to-STAR-RIS users may be divided into the same group. This is adverse to enhancing the total sum rate, because these users would compete for the same-side beamforming of the STAR-RIS. Hence, the number of users on both sides of the STAR-RIS should be balanced as well. In this way, the two-side beamforming vectors of STAR-RIS can be fully used to boost the network performance.   

Based on above analysis, we define a differential distance to measure the sum-rate gap of the two-side users
\begin{equation} 	\label{eq-d-both sides}
			\setlength{\abovedisplayskip}{3pt}
	\setlength{\belowdisplayskip}{3pt}
	D_c= |\bar{d}^\ft_c-\bar{d}^\fr_c|\times\max(|N^\ft_c-N^\fr_c|, 10^{-3}),
\end{equation}
where $c\in\mathcal{K}$ is the candidate boundary user index; $\bar{d}^\ft_c$ and $\bar{d}^\fr_c$ are the average distances of STAR-RIS-to-transmission-users and STAR-RIS-to-reflection-users, respectively; $N^\ft_c$ and $N^\fr_c$ are the numbers of transmission- and reflection- users, respectively. By minimizing (\ref{eq-d-both sides}), the sum rate of users at both sides of the STAR-RIS can be well balanced.  

With the measure $D_c$, we construct a differential distance-based roulette wheel to determine the boundary user. The roulette wheel is one-armed, where the size of the holes reflects the selection probabilities of each user being the boundary user. The user with a small differential distance is given a high selection probability, which will be prioritised as the boundary user. The whole process of the balanced grouping method is given in Algorithm \ref{al-RW-grouping}.

\begin{algorithm}[t]
	\caption{SDP method for hybrid beamforming design.} 
	\begin{algorithmic}[1]
		\State Initialize $\wt_k$, $\mta_\ft$, and $\mta_\fr$.
		\Statex \textbf{REPEAT}
		\State Update $\wt_k$, $A_k$ and $B_k$ by solving the relaxed version of problem (\ref{pb-BS-beamforming}) with CVX toolbox; 
		\State Update $\mta_\ft$, $\mta_\fr$, $A_k$ and $B_k$ by solving the relaxed version of problem (\ref{pb-STAR-beamforming}) with CVX toolbox;
		\Statex \textbf{Until} the achievable sum-rate converges. 					
	\end{algorithmic}	\label{al-lower level}
\end{algorithm}

\subsubsection{SDP Method} \label{sec-lower level}
With given STAR-RIS deployment, problem (\ref{pb-SR2}) is reduced to a hybrid beamforming design problem:
\begin{subequations}\label{pb-hybrid-beamforming}
	\setlength{\abovedisplayskip}{3pt}
	\setlength{\belowdisplayskip}{3pt}
	\begin{align}
		\max_{\wt_k, \mta_\ft, \mta_\fr}\quad
		&\sum_{k=1}^{K}R_k, \tag{\ref{pb-hybrid-beamforming}{a}}\label{pb-hybrid-beamforminga}\\ 
		s.t.\quad\ &(\ref{pb-SR2d}).  \tag{\ref{pb-hybrid-beamforming}{b}}\label{pb-hybrid-beamformingb} 
	\end{align}	
\end{subequations}
We decouple this problem into two subproblems: active beamforming at BS and passive beamforming at STAR-RIS. Here we employ the SDP method in \cite{gao2022joint} for hybrid beamforming. The pseudocode is given in Algorithm \ref{al-lower level}.

\textbf{Active Beamforming Optimization:} According to the SDP method \cite{gao2022joint}, we define a slack variable set $\{A_k, B_k\}$ 
\begin{equation}\label{eq-ABk}
	\setlength{\abovedisplayskip}{3pt}
\setlength{\belowdisplayskip}{3pt}
	\frac{1}{A_k}=|\hht_k\wt_k|^2,\quad B_k=\sum_{j\neq k, j\in\mathcal{K}}|\hht_k\wt_j|^2+\sigma_0^2.
\end{equation}
By substituting (\ref{eq-ABk}) into problem (\ref{pb-hybrid-beamforming}), denoting $\Ht_k=\hht_k^\fH\hht_k$, and $\Wt_k=\wt_k\wt_k^\fH$, and assuming that the STAR-RIS beamforming is given, problem (\ref{pb-hybrid-beamforming}) is reduced to an active beamforming subproblem 
\begin{subequations}\label{pb-BS-beamforming}
		\setlength{\abovedisplayskip}{3pt}
	\setlength{\belowdisplayskip}{3pt}
	\begin{align}
		\max_{\Wt_k, A_k, B_k, R_k}\quad
		&\sum_{k=1}^{K}R_k, \tag{\ref{pb-BS-beamforming}{a}}\label{pb-BS-beamforminga}\\ 
		s.t.\quad\ &\log_2\left(1+\frac{1}{A_kB_k}\right)\geq R_k, \tag{\ref{pb-BS-beamforming}{b}}\label{pb-BS-beamformingb}\\ 
		&\frac{1}{A_k}\leq \text{Tr}(\Ht_k\Wt_k),\  \tag{\ref{pb-BS-beamforming}{c}}\label{pb-BS-beamformingc}\\	
		&B_k\geq\sum_{j\neq k, j\in\mathcal{K}}\text{Tr}(\Ht_k\Wt_j)+\sigma_0^2,\  \tag{\ref{pb-BS-beamforming}{d}}\label{pb-BS-beamformingd}\\
		&\sum_{k=1}^{K}\text{Tr}(\Wt_k)\leq P_{max},  \tag{\ref{pb-BS-beamforming}{e}}\label{pb-BS-beamforminge} \\
		&\text{Rank}(\Wt_k)=1,\ \tag{\ref{pb-BS-beamforming}{f}}\label{pb-BS-beamformingf}\\ 
		&\Wt_k\succeq0,\ \tag{\ref{pb-BS-beamforming}{g}}\label{pb-BS-beamformingg} \\
		&(\ref{pb-SRd}), \tag{\ref{pb-BS-beamforming}{h}}\label{pb-BS-beamformingh}
	\end{align}
\end{subequations}
To handle the non-convex constraints (\ref{pb-BS-beamformingb}), the first-order Taylor expansion is employed to attain its lower bound \cite{9139273}:
\begin{equation} \label{eq-AB-Taylor}
		\setlength{\abovedisplayskip}{3pt}
	\setlength{\belowdisplayskip}{3pt}
	\begin{aligned}
		\log_2(1+\frac{1}{A_kB_k})\geq \widehat{R}_k 
		\triangleq \log_2(1+\frac{1}{\hat{A}_k^{l_1}\hat{B}_k^{l_1}})-\frac{\log_2 e(A_k-\hat{A}_k^{l_1})}{\hat{A}_k^{l_1}(\hat{B}_k^{l_1}\hat{B}_k^{l_1}+1)}-\frac{\log_2 e(B_k-\hat{B}_k^{l_1})}{\hat{A}_k^{l_1}(\hat{A}_k^{l_1}\hat{B}_k^{l_1}+1)},
	\end{aligned}
\end{equation}
where $\hat{A}_k^{l_1}$ and $\hat{B}_k^{l_1}$ are the values of $A_k$ and $B_k$ in the $l_1$-th iteration, respectively. The rank-1 constraint (\ref{pb-BS-beamformingf}) is provably satisfied \cite{9570143}. Therefore, by replacing (\ref{pb-BS-beamformingb}) with $\widehat{R}_k\geq R_k$, and drop the rank-1 constraint, we obtain the relaxed version of Problem (\ref{pb-BS-beamforming}) and solve it by the CVX toolbox \cite{cvx}.    

\textbf{Transmission- and Reflection- Beamforming Optimization:} Inspired by the SDP method in \cite{gao2022joint}, we define $\widehat{\hht}_k=$\text{diag}$(\vt_k^\fH)\GGt\wt_k$, $\widehat{\Ht}_k=\widehat{\hht}_k^\fH\widehat{\hht}_k$, $\widehat{\mTa}_\varrho=\mta_\varrho\mta_\varrho^\fH$, where $\varrho\in\{\ft,\fr\}$, substitute them into (\ref{eq-ABk}), then substitute (\ref{eq-ABk}) into Problem (\ref{pb-hybrid-beamforming}), use the relaxed constraint $\widehat{R}_k\geq R_k$ instead of (\ref{pb-BS-beamformingb}), Problem (\ref{pb-hybrid-beamforming}) is reduced to a transmission- and reflection- beamforming problem
\begin{subequations}\label{pb-STAR-beamforming}
		\setlength{\abovedisplayskip}{3pt}
	\setlength{\belowdisplayskip}{3pt}
	\begin{align}
		\max_{\widehat{\mTa}_\varrho, A_k, B_k, R_k}\quad
		&\sum_{k=1}^{K}R_k, \tag{\ref{pb-STAR-beamforming}{a}}\label{pb-STAR-beamforminga}\\ 
		s.t. \quad\ &\widehat{R}_k\geq R_k, \\
 &\frac{1}{A_k}\leq \text{Tr}(\widehat{\mTa}_\varrho\widehat{\Ht}_k),\  \tag{\ref{pb-STAR-beamforming}{b}}\label{pb-STAR-beamformingb}\\	
		&B_k\geq\sum_{j\neq k, j\in\mathcal{K}}\text{Tr}(\widehat{\mTa}_\varrho\widehat{\Ht}_j)+\sigma_0^2,\  \tag{\ref{pb-STAR-beamforming}{c}}\label{pb-STAR-beamformingc}\\
		&\widehat{\mTa}_\varrho\succeq0,\ \tag{\ref{pb-STAR-beamforming}{d}}\label{pb-STAR-beamformingd}\\		
		&\text{Rank}(\widehat{\mTa}_\varrho)=1,\ \tag{\ref{pb-STAR-beamforming}{e}}\label{pb-STAR-beamforminge}\\ 
		&(\ref{pb-SRd}), (\ref{pb-SRe}), (\ref{pb-SRf}), (\ref{pb-BS-beamformingb}),  \tag{\ref{pb-STAR-beamforming}{d}}\label{pb-STAR-beamformingf}   
	\end{align}	
\end{subequations}

The rank-1 constraint (\ref{pb-STAR-beamforminge}) is non-convex, which can be relaxed by \cite{liu2021star}
\begin{equation}
	\xi_\text{max}(\widehat{\mTa}_\varrho)\geq\et^\fH_{max}(\widehat{\mTa}_\varrho^{l_2})\widehat{\mTa}_\varrho \et_{max}(\widehat{\mTa}_\varrho^{l_2})\overset{(a)}{\geq}\varepsilon^{l_2}\text{Tr}(\widehat{\mTa}_\varrho),\  \varrho\in\{\ft,\fr\},
	\label{eq-rank1-relax}
\end{equation}
where $\xi_\text{max}(\widehat{\mTa}_\varrho)$ is the largest eigenvalue of $\widehat{\mTa}_\varrho$. $\et^\fH_{max}(\widehat{\mTa}_\varrho^{l_2})$ is the eigenvector corresponding to the maximum eigenvalue of $\widehat{\mTa}_\varrho^{l_2}$ at the ${l_2}$-th generation. By sequentially increasing the parameter $\varepsilon^{l_2}$ over iterations, the rank-1 solution is gradually attained. Thus, by replacing \eqref{pb-STAR-beamforminge} with inequation (a), we obtain the relaxed version of problem 
(\ref{pb-STAR-beamforming}) and solve it by CVX toolbox \cite{cvx}.

\subsubsection{Deployment Orientation Refinement}
Fixing the hybrid beamforming, Problem (\ref{pb-SR2}) becomes
\begin{subequations}\label{pb-ori}
		\setlength{\abovedisplayskip}{3pt}
	\setlength{\belowdisplayskip}{3pt}
	\begin{align}
		\max_{\phi}\
		&\sum_{k=1}^{K}R_k, \tag{\ref{pb-ori}{a}}\label{pb-oria}\\ 
		s.t.\ & \phi\in[\phi_{(c)}, \phi_{(c+1)})+0.001,\ (\ref{pb-SRd}). \tag{\ref{pb-ori}{b}}\label{pb-orib} 	
	\end{align}
\end{subequations}
Optimizing this problem is very challenging due to multiple cosine and sine components in \eqref{eq-array vector1}. To address this issue, we propose a gridding method. Specifically, we partition the on-hand angular region of $\phi$ into several grids with the interval $\triangle\leq \triangle_{max}$, where $\triangle_{max}$ is the predefined threshold. Following the feasible solution technique \cite{7277038}, the best orientation is the grid point with the highest sum rate and lowest QoS violation. Considering the performance improvement and complexity, we only perform the orientation refinement once for each individual. 

\subsection{Convergence and Complexity Analysis}\label{sec-convergence}
It is challenging to verify mathematically the convergence of the DE reproduction and balanced grouping. Thus, we use simulations for validation. We can prove the convergence of the SDP method according to \cite{gao2022joint}. In the orientation refinement, the gridding method is a greedy search method, the convergence always holds. Despite that, the DE selection operator is an elitism selection operator. It follows the feasible solution technique, always prioritising a solution with a high sum rate value and low QoS violation. Thus the best sum-rate value is non-increasing after each generation. In addition, since the sum rate is upper bound by a finite value, the proposed algorithm is guaranteed to converge. 

The complexity mainly depends on the hybrid beamforming. Specifically, the complexity of solving the active beamforming optimization problem (\ref{pb-BS-beamforming}) is $O_1\triangleq\mathcal{O}\left(l_1^{max}\text{max}(N_a, (2K+1))^4\sqrt{N_a}\log_2\frac{1}{\epsilon_1}\right)$, where $l_1^{max}$ is the maximum iteration number, and $\epsilon_1$ is the solution accuracy. The complexity of solving the STAR-RIS's transmission- and reflection- beamforming problem (\ref{pb-STAR-beamforming}) is $O_2\triangleq\mathcal{O}\left(l_2^{max}\text{max}(M, 2K)^4\sqrt{M}\log_2\frac{1}{\epsilon_2}\right)$, where $l_2^{max}$ is the maximum iteration number, and $\epsilon_2$ is the solution accuracy. Therefore, the total complexity is $\mathcal{O}\left(NG^{max}l_3^{max}(O_1+O_2)\right)$, where $N$ is the population size, $G^{max}$ and $l_3^{max}$ are the maximum generation for outer loop and hybrid beamforming, respectively. 

\section{Simulation Results}
In this section, numerical simulations are conducted to evaluate the performance of DEBG.

\subsection{Setup, Metrics and Algorithms} 
The STAR-RIS-assisted communication system in Fig. \ref{fig-system} is investigated. The BS is located at [0,0,0], and the STAR-RIS is deployed within the region $\{25\leq x_s\leq80, -10\leq y_s\leq20, z_s=2\}$. Assume that there are 2 clusters in the network and each cluster contains 4 users, where the cluster centres are [30,-5,0] and [65,5,0], with a radius of 5 m and 15 m, respectively. The reference path loss at 1 meter is $\rho_0=-30$ dB. All links' path loss exponents and Rician factors are $\alpha=2.2$ and $\beta_{BS}=\beta_{SU_k}=3$ dB. The noise power is $\sigma_0^2=-90$ dBm. For each user, the QoS requirement is $R_{min}=0.1$ bits/s/Hz. The number of STAR-RIS elements is $M=M_hM_v$, where $M_v$ and $M_h$ are the number of elements along the x-y plane and z-axis, respectively. Here $M_h$ is set to 5. The element spacing of the STAR-RIS and BS is set to $d_I=d_L=\lambda/2$. Unless stated otherwise, the number of BS antennas equals $N_a=4$, the number of STAR-RIS elements is $M=20$, and the maximum transmit power at the BS is $P_{max}=20$ dBm. 

To validate the effectiveness of the proposed DEBG approach, we compare it with the following baseline algorithms: 1) \textit{RandBeamforming}: The hybrid beamforming is randomly designed. 2) \textit{RandDeploy}: The location and orientation of the STAR-RIS are both randomly generated. 3) \textit{RandLoc}: The STAR-RIS's location in this algorithm is randomly decided. 4) \textit{AO-FixedGrouping}: A state-of-the-art work \cite{gao2022joint} with the assumption of fixed user grouping. Specifically, an initial user grouping is randomly generated and remains unchanged, then the alternating optimization (AO) method is evoked to alternately optimize the STAR-RIS's location, the active and passive beamforming. 5) \textit{RandGrouping}: The boundary user is randomly determined such that the user grouping is randomly decided. 

For our DEBG approach and its variants, the population size is set to $N=20$. The amplification factor and crossover rate are set to $F=0.5$ and $\text{Cr}=0.4$, based on simulation results in supplement file. The threshold for orientation interval is set to $\triangle=0.1$. The sum rate results are obtained by averaging over 50 user distributions for a fair comparison. Specifically, the corresponding channel realization is generated once for each user distribution, and each algorithm runs 30 times. The tolerance convergence for hybrid beamforming is set to $\epsilon_1, \epsilon_2=10^{-4}$. All the algorithms terminate if the number of explored STAR-RIS deployment exceed 1200. That is, for RandGrouping and DEBG, the maximization generation $G$ is 60 due to the population size $N=20$. We used the Wilcoxon rank-sum test \cite{demvsar2006statistical} with a 0.05 significance level to statistically compare the sum rate of different algorithms for each problem instance. 

\subsection{Comparison of DEBG Against Other Methods}

\subsubsection{Performance versus the number of STAR-RIS elements}
Table \ref{tab-reM} presents the mean and standard deviation of sum rate values (bits/s/Hz) with respect to $M$. It is evident that as $M$ increases, all methods achieve higher sum rates. This observation suggests that a larger number of STAR-RIS elements offer DoFs, aiding in balancing signal power and mitigating interference from other users. As anticipated, methods such as RandLoc, AO-FixedGrouping, and RandGrouping outperform the RandDeploy approach, attributed to their optimization of either the location or orientation of the STAR-RIS. Notably, our DEBG method demonstrates superior performance, followed by RandGrouping. This demonstrates the significance of optimizing deployment orientation in enhancing overall performance.

\textbf{\begin{table*}[t]
	\centering
	\caption{Mean and standard deviation (std) of sum rate results (bits/s/Hz) over different numbers of STAR-RIS elements $M$ with $P_{max}=20$ dBm and $N_a=4$.}
	\label{tab-reM}
	\centering
  \resizebox{\textwidth}{!}{
	\begin{tabular}{cccccccc}
		\hline
		$M$ & &RandBeamforming & RandDeploy         & RandLoc       &  AO-FixedGrouping          &  RandGrouping                 & DEBG \\ \hline
        \multirow{2}{*}{20} & mean & 16.41 & 49.87 & 53.49 & 64.79 & 66.47  & \textbf{72.95} \\
                            & std  & 5.25  & 3.15  & 2.89 & 2.76  & 2.05    & 1.34 \\ \hline
        \multirow{2}{*}{30} & mean & 16.16 & 52.09 & 56.06 & 66.22 & 68.66  & \textbf{74.01} \\
                            & std  & 4.89  & 3.14  & 2.84 & 2.67  & 2.18    & 1.63 \\ \hline                 
        \multirow{2}{*}{40} & mean & 17.26 & 53.43 & 58.64 & 68.25 & 70.58  & \textbf{75.69} \\
                            & std  & 6.42  & 3.23  & 2.79 & 2.63  & 2.06    & 1.92 \\ \hline                 
        \multirow{2}{*}{50} & mean & 18.70 & 55.79 & 60.58 & 70.04 & 72.29 & \textbf{77.81} \\
                            & std  & 6.78  & 3.09   & 2.75  & 1.97  &2.79  & 1.85 \\  \hline                
        \multirow{2}{*}{60} & mean & 19.04 & 57.09 & 62.17 & 72.96 & 74.16  & \textbf{79.06} \\
                            & std  & 7.12  & 3.33  & 2.81 & 2.86  & 1.99   & 1.70 \\ \hline		
	\end{tabular}
 }
	\end{table*}  }
	
\subsubsection{Performance versus the number of BS antennas}
Table \ref{tab-reNa} gives the sum rate results (bits/s/Hz) over $N_a$, where the transmit power budget of BS is $P_{max}=20$ dBm, and the number of STAR-RIS elements is $M=20$. It can be seen that all the methods obtain higher sum rate values as the number of BS antenna grows. This is because increasing antennas enables a higher active beamforming gain. The RandGrouping and AO-FixedGrouping perform better than RandDeploy, because they employ DE or AO to achieve a high-quality deployment location for STAR-RIS. Compared to other algorithms, a considerable performance enhancement is observed from the proposed DEBG, demonstrating the importance of joint optimization of STAR-RIS's location and user grouping. 
	
\begin{table*}[t]
\renewcommand{\arraystretch}{1}
\centering
\caption{Mean and standard deviation (std) of sum rate results (bits/s/Hz) over different numbers of BS antennas $N_a$ with $P_{max}=20$ dBm and $M=20$.}
\label{tab-reNa}
\centering
 \resizebox{\textwidth}{!}{
\begin{tabular}{cccccccc}
	\hline
	$N_a$  & &RandBeamforming  & RandDeploy & RandLoc  &  AO-FixedGrouping  &  RandGrouping            & DEBG \\ \hline
        \multirow{2}{*}{4} & mean & 16.41 & 49.87 & 53.49 & 64.79 & 66.47  & \textbf{72.95} \\
                            & std  & 5.25  & 3.15  & 2.89 & 2.76  & 2.05    & 1.34 \\ \hline
                            
        \multirow{2}{*}{5} & mean & 17.81 & 51.69 & 55.25 & 66.58 & 68.64  & \textbf{73.84} \\
                            & std  & 7.32  & 3.42 & 2.89 & 2.67  & 2.13    & 1.76 \\ \hline     
                            
        \multirow{2}{*}{6} & mean & 18.59 & 52.40  & 58.40 & 68.87 & 70.55  & \textbf{75.07} \\
                            & std  & 6.21  & 2.98  & 2.88 & 2.62  & 1.82    & 2.06 \\ \hline                 
        \multirow{2}{*}{7} & mean & 19.04 & 54.08 & 60.97 & 70.95 & 72.63 & \textbf{76.12} \\
                            & std  & 6.73  & 2.65   & 2.86  & 2.71  & 1.67  & 1.68 \\  \hline     
                            
        \multirow{2}{*}{8} & mean & 20.31 & 56.38 & 63.05 & 72.77 & 74.26  & \textbf{77.64} \\
                            & std  & 6.96  & 2.80  & 2.67 & 2.66  & 1.67   & 1.89 \\ \hline			
\end{tabular}%
} 
\end{table*}

\subsubsection{Final STAR-RIS deployment result of DEBG}
\begin{figure}[t] 
\centering
\includegraphics[width=0.4\textwidth]{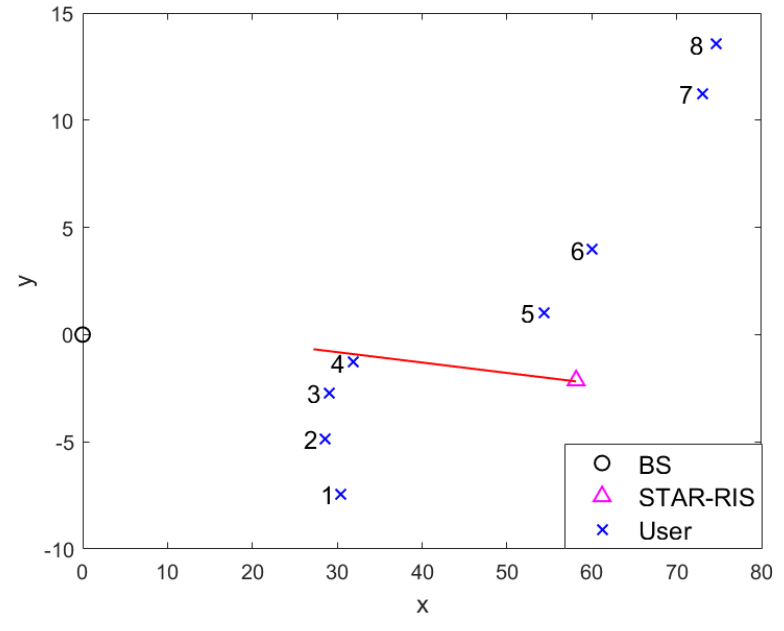}
\caption{Optimal STAR-RIS deployment (top view).} 
\label{fig-final-deploy}
\end{figure}

Fig. \ref{fig-final-deploy} shows the optimal STAR-RIS deployment obtained by DEBG under $P_{max}=20$, $N_a=4$, and $M=40$. The optimal STAR-RIS deployment location is $(56.43,-1.55)$ with an orientation $178.43^\circ$. As can be observed, the STAR-RIS deployment is more symmetric across all users and nearer the users. This is because the received signal power in the STAR-RIS assisted link generally scales with $1/(d_{BS}^\alpha d_{SU}^\alpha)$. The channel gain of user $k$ increases when the STAR-RIS gets close and decreases when it gets away. Therefore, the STAR-RIS is deployed near all the users rather than the BS. In addition, the STAR-RIS tends to divide STAR-RIS-nearer users with higher channel gains. Such a setting enables a different STAR-RIS beamforming allocation for these users as far as possible to get the most performance out of it. Following this setting, the users are divided into two groups: users $1\sim5$ and $6\sim8$. However, in the first group, four users (users $1\sim4$) sacrifice their performance to maximize the performance of user 5, and suppress the strong interference from users 5 and 6. However, the sacrifice is too huge to complement the performance improvement of users 5 and 6, which is adverse to maximizing the whole performance. To alleviate the sacrifice, user 5 is finally moved to the second group such that the performance of two groups can be well balanced. The grouping result demonstrates the superiority of the proposed balanced grouping method.

\subsubsection{Robustness Analysis to Imperfect CSI}
\begin{figure}[t] 
	\centering
	\includegraphics[width=0.37\textwidth]{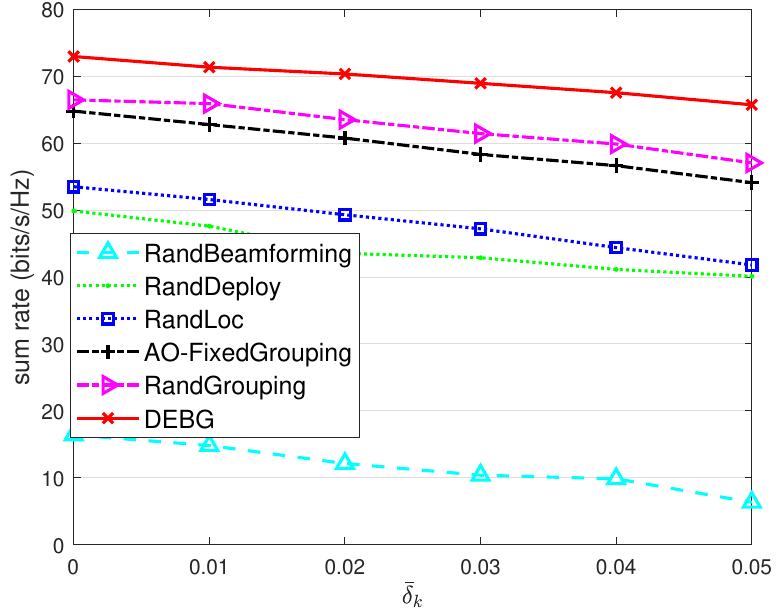}
	\caption{Average worst-case sum rate versus the uncertainty level of cascaded channels under $P_{max}=20 dBm$, $N_a=4$ and $M=20$.} 
	\label{fig-imperfect}
\end{figure}

Our approach optimises the deployment and hybrid beamforming based on the deterministic LOS components. The CSI is imperfect due to the ignorance of random NLOS components. To investigate the impact of the imperfect CSI, we adopt a bounded model to characterize the CSI imperfection. Specifically, the imperfect cascaded model is modelled as $\hht_k=\widehat{\hht}_k+\triangle\hht_k$, where $\widehat{\hht}_k$ is the estimate of $\hht_k$, and $\triangle\hht_k$ is the unknown CSI error limited in the regions of constant radii $\delta_k$. For simplicity, we define $\bar{\delta}_k=\|\triangle\hht_k\|_F/\|\hht_k\|_F, \forall k$. Fig. \ref{fig-imperfect} illustrates the average worst-case sum rate versus the uncertainty level of cascaded channels $\bar{\delta_k}$. As the uncertainty level $\bar{\delta_k}$ increases from 0 to 0.05, the worst-case sum rate of all algorithms decreases sharply. This is because the CSI error of the cascaded channel incurs a diversity gain and brings a severe performance loss. Despite that, DEBG shows the best robustness under all uncertainty levels.
	
\subsection{Detailed Analysis of DEBG} 
This subsection investigates the convergence behaviour and the balanced grouping method to demonstrate the superiority of DEBG. Finally, some discussions are presented. In addition, the effectiveness of the DE operators and parameter sensitivity are investigated in the supplementary file.

\subsubsection{Convergence behavior}
\begin{figure*}[t] 
\centering
\subfigure[]{\includegraphics[width=0.32\textwidth,height=0.25\textwidth]{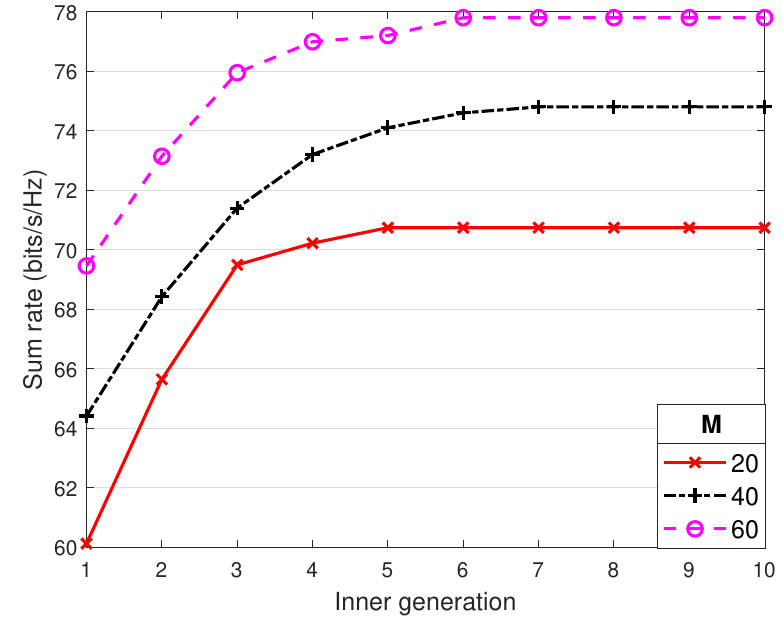}} 
\subfigure[]{\includegraphics[width=0.32\textwidth,height=0.25\textwidth]{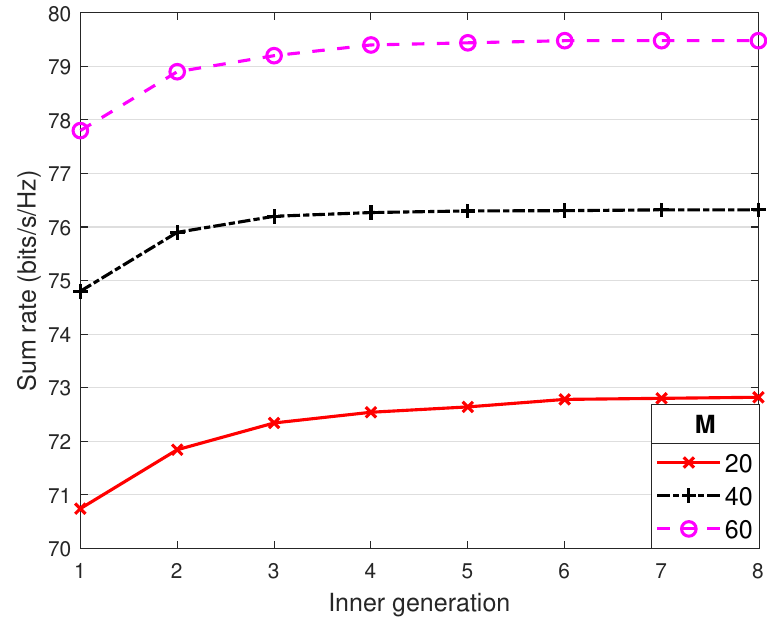}} 	\subfigure[]{\includegraphics[width=0.32\textwidth,height=0.25\textwidth]{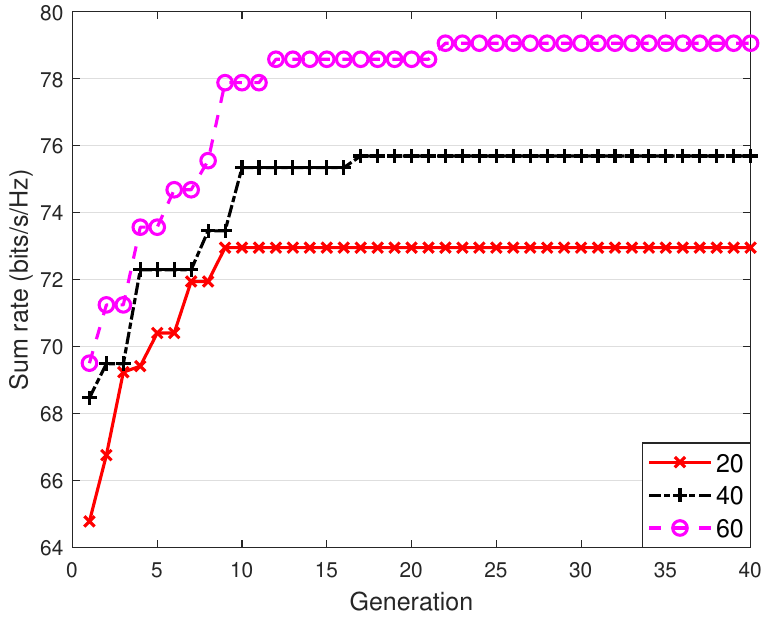}}
\caption{Convergence behaviour of (a) hybrid beamforming, (b) orientation refinement, and (c) location optimization and balanced grouping, where $P_{max}=20$ dBm and $N_a=4$.} 
\label{fig-DEBG-convegence}
\end{figure*}

Fig. \ref{fig-DEBG-convegence} illustrates the typical convergence behaviour of DEBG components, where $P_{max}=20$ dBm, $N_a=4$, and $M\in\{20,40,60\}$. The maximum generation $G^{max}$ is set to 50. In Fig. \ref{fig-DEBG-convegence}(a), the hybrid beamforming optimization algorithm converges within 10 iterations under different $M$s. From Fig. \ref{fig-DEBG-convegence}(b), it can be found that the orientation refinement needs at most 8 iterations to converge. The location and boundary-user index quality cannot be independently evaluated but only estimated with corresponding low-level variables (i.e., hybrid beamforming and orientation). Therefore, Fig. \ref{fig-DEBG-convegence} (c) presents the collaborative convergence of location optimization and balanced grouping. Fig. \ref{fig-DEBG-convegence} (c) shows that as the number of STAR-RIS elements grows, DEBG performs better and requires more converging iterations. Note that DEBG achieves the best performance within 25 generations for the considered cases. 

\subsubsection{Study of balanced grouping method}
To validate the effectiveness of balanced grouping, we consider three DEBG variants varying from user grouping strategies: i) OneGroup strategy. The STAR-RIS is replaced by a reflection-only RIS, where this RIS also has $M$ elements and is deployed parallel to the z-axis. This is a special case of STAR-RIS, i.e., all the users are in the same group. ii) Rand\_$\phi$ strategy. The STAR-RIS deployment is represented by STAR-RIS's location and orientation $\phi$. The orientation is randomly selected within $[0,2\pi)$. iii) Rand\_$c$ strategy. The proposed point-point representation characterizes the STAR-RIS deployment, in which the boundary user index $c$ is randomly chosen from $\mathcal{K}$. 

\begin{figure}[t] 
\centering
\includegraphics[width=0.37\textwidth]{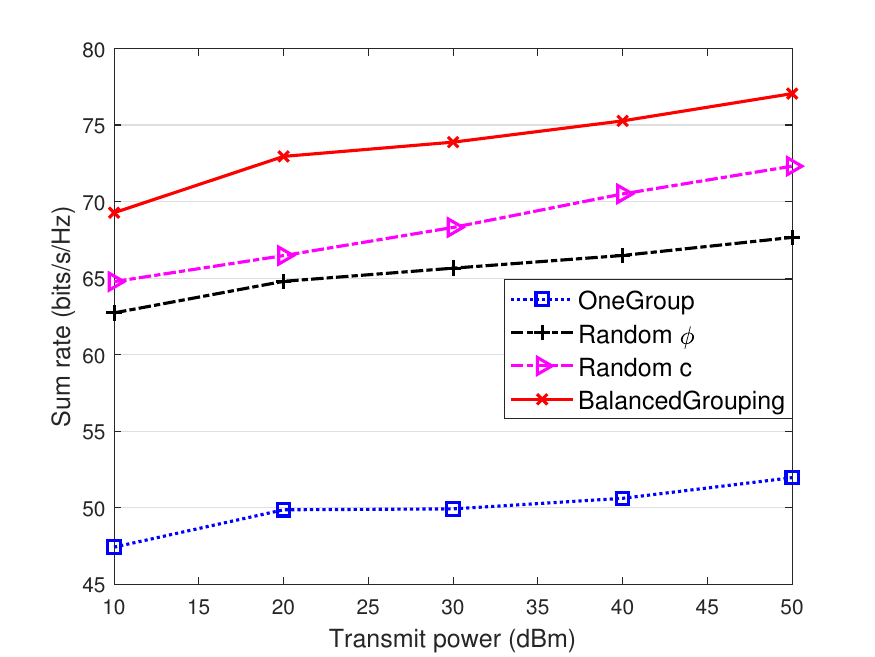}
\caption{Performance comparison of different grouping strategies in DEBG with $N_a=4$ and $M=20$.} 
\label{fig-reGrouping}
\end{figure}

Fig. \ref{fig-reGrouping} illustrates the performance comparison of different grouping strategies versus the transmit power $P_{max}$. Rand\_$\phi$ and Rand\_$c$ achieve higher sum rates than OneGroup strategy. This advantage benefits from the more DoFs for simultaneous transmission- and reflection- beamforming. Compared to the Rand\_$\phi$ strategy, Rand\_$c$ strategy obtains better performance, since the boundary user possesses more compact search space and can improve the search efficiency of user grouping. Our balanced grouping performs the best in all problem cases, demonstrating its effectiveness in user grouping. 

\subsection{Discussions}
\textbf{Effect of the initial value and step size of STAR-RIS location:} Unlike \cite{9852464}, our approach is unaffected by the location's initial value and step size. Multiple solutions are generated at the initialization to avoid it being affected by the initial value effectively. In addition, the step size of the location can be measured by the mutation operator \eqref{eq-mutation}. The step size can be seen as adaptive and decreasing towards zero with iterations. Specifically, at early iterations, the locations of solutions are relatively dispersed and optimized in a wide search space, where the step size is ``large ''. As the iteration progresses, the solutions converge towards the optimum and get much closer to each other. The step size gets ``smaller ''. Until convergence, the step size would be approximately zero.

\textbf{Applicability:} The proposed approach is particularly well-suitable for attaining a satisfying deployment scheme for the STAR-RIS in quasi-static scenarios, such as outdoor temporary hotspots and IoT networks. While for scenarios with significant randomization of NLOS components or dynamic user mobility, the optimal deployment strategy of STAR-RIS will be explored in future research. In addition, when considering the hardware implementation, the STAR-RIS beamforming should obey the correlated transmission and reflection phase-shift model. Our method can be extended to achieve this by incorporating the phase-shift configuration strategies in \cite{9774942}.

\section{Conclusion}
This paper addressed the deployment location and orientation problem in a STAR-RIS-assisted network. We formulated the joint STAR-RIS deployment, active and passive beamforming optimization problem for the sum-rate maximization, subject to the transmit power budget at the BS and the QoS constraint for each user. To solve this problem, we proposed the point-point representation to characterize the user grouping concisely, providing a solid problem-solving basis. Then, a differential evolution approach with balanced grouping was developed as the solver. In particular, the balanced grouping method enables determining a matching boundary user for a given STAR-RIS deployment location, so the desired user grouping can be achieved. Our simulation results validated the significant performance enhancement of the proposed STAR-RIS deployment design. A more symmetric STAR-RIS deployment strategy is preferable, which provides useful guidance for practical STAR-RIS applications.

\section*{Acknowledgment}
This work is partly supported by the National Natural Science Foundation of China under Grant No. 61701216, Shenzhen Science, Technology, and Innovation Commission Basic Research Project under Grant No. JCYJ20180507181527806, Guangdong Provincial Key Laboratory under Grant No.2020B121201001, Guangdong Innovative and Entrepreneurial Research Team Program under Grant No.2016ZT06G587, and Shenzhen Sci-Tech Fund under Grant No. KYTDPT20181011104007. 

\bibliographystyle{elsarticle-num} 
\bibliography{Ref_STAR}

\end{document}